# Quantum state evolution in $C^2$ and $G_3^+$

*Alexander M. SOIGUINE*

**Abstract:** It was shown, [1] - [3], that quantum mechanical qubit states as elements of $C^2$ can be generalized to elements of even subalgebra $G_3^+$ of geometric algebra $G_3$ over Euclidian space $E_3$. The construction critically depends on generalization of formal, unspecified, complex plane to arbitrary variable, but explicitly defined, planes in 3D, and of usual Hopf fibration $S^3 \to S^2$ to maps $\{g : g \in G_3^+, |g| = 1\} \xrightarrow{B} S^2$ generated by arbitrary unit value bivectors $B \in G_3^+$. Analysis of the structure of the map $G_3^+ \to C^2$ demonstrates that quantum state evolution in $C^2$ gives only restricted information compared to that in $G_3^+$.

1. Introduction

Qubits, unit value elements of the Hilbert space $C^2$ of two dimensional complex vectors, can be generalized to unit value elements $\alpha + I_S \beta$ [1] of even subalgebra $G_3^+$ of geometric algebra $G_3$ over Euclidian space $E_3$. I called such generalized qubits $g$-**qubits** [1].

Some minimal information about $G_3$ and $G_3^+$ is necessary. Algebraically, $G_3$ is linear space with canonical basis $\{1, e_1, e_2, e_3, e_2e_3, e_3e_1, e_1e_2, e_1e_2e_3\}$, where 1 is unit value scalar, $\{e_i\}$ are orthonormal basis vectors in $E_3$, $\{e_i e_j\}$ are oriented, mutually orthogonal unit value areas (bivectors) spanned by $e_i$ and $e_j$ as edges, with orientation defined by rotation $e_i$ to $e_j$ by angle $\frac{\pi}{2}$; and $e_1 e_2 e_3$ is unit value oriented volume spanned by ordered edges $e_1$, $e_2$ and $e_3$.

Subalgebra $G_3^+$ is spanned by scalar 1 and basis bivectors: $\{1, e_2e_3, e_3e_1, e_1e_2\}$. Variables $\alpha$ and $\beta$ in $\alpha + \beta I_S \in G_3^+$ are scalars, $I_S$ is a unit size oriented area, bivector, lying in an arbitrary given plane $S \subset E_3$. Bivector $I_S$ is linear combination of basis bivectors $e_i e_j$, see Fig.1a.

---

[1] This sum bears the sense *"something and something"*. It is not a sum of geometrically similar elements giving another element of the same type, see [3], [9].

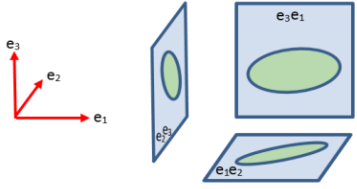 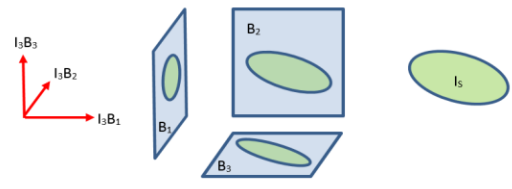

Fig.1a                            Fig.1b

It was explained in [4], [5] that elements $\alpha + I_S \beta$ only differ from what is traditionally called "complex numbers" by the fact that $S \subset E_3$ is an arbitrary, though explicitly defined, plane in $E_3$. Putting it simply, $\alpha + I_S \beta$ are "complex numbers" depending on $E_2$ embedded into $E_3$. Traditional "imaginary unit" $i$ is actually associated with some unspecified $I_S$ – everything is going on in one tacit fixed plane, not in 3D world. Usually $i$ is considered just as a "number" with additional algebraic property $i^2 = -1$. That may be a source of ambiguities, as it happens in quantum mechanics.

I will denote unit value oriented volume $e_1 e_2 e_3$ by $I_3$. It has the property $I_3^2 = -1$. Actually, there always are two options to create oriented unit volume, depending on the order of basis vectors in the product. They correspond to the two types of handedness – left and right screw handedness. In Fig.1a the variant of right screw handedness is shown. It is geometrically obvious that $e_i e_j = -e_j e_i$ (bivector flips, changes orientation to the opposite, in swapping the edge vectors). Then changing of handedness, for example by $e_1 e_2 e_3 \to e_2 e_1 e_3$, gives $e_2 e_1 e_3 = -e_1 e_2 e_3 = -I_3$.

Any basis vector is conjugate (conjugation means multiplication by $I_3$), or dual, to a basis bivector and inverse. For example $I_3 e_2 e_3 = e_2 e_3 I_3 = -e_1$ . By multiplying these equations from left or right by $I_3$ we get $e_2 e_3 = I_3 e_1 = e_1 I_3$. Such duality holds for arbitrary, not just basis, vectors and bivectors. Any vector $\vec{b}$ is conjugate to some bivector: $\vec{b} = I_3 B = B I_3$ and for any bivector $B$ there exists $\vec{b}$ such that $B = I_3 \vec{b} = \vec{b} I_3$. Good to remember that in $\vec{b} = I_3 B = B I_3$ vector and bivector are in handedness opposite to $I_3$ while in $B = I_3 \vec{b} = \vec{b} I_3$ they are in the same handedness as $I_3$.

One can also think about $I_3$ as a right (left) single thread screw helix of the height one. In this way $-I_3$ is left (right) screw helix.

All above remains true (see Fig.1b) if we replace basis bivectors $\{e_2 e_3, e_3 e_1, e_1 e_2\}$ by an arbitrary triple of unit bivectors $\{B_1, B_2, B_3\}$ satisfying the same multiplication rules which are valid for $\{e_2 e_3, e_3 e_1, e_1 e_2\}$ (see Fig.2, right screw handedness assumed):

$$B_1 B_2 = -B_3; \quad B_1 B_3 = B_2; \quad B_2 B_3 = -B_1 \qquad (1.1)$$

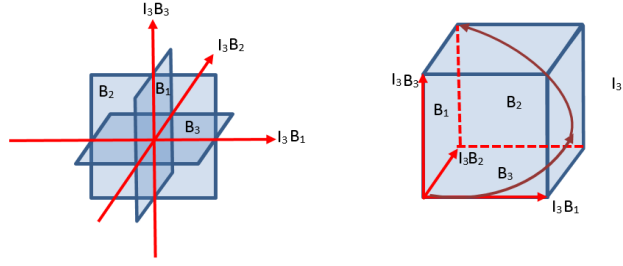

Fig.2

## 2. Parameterization of unit value elements in $G_3^+$ and $C^2$ by points of $S^3$

Let's take a $g$-qubit:

$$\alpha + I_S\beta = \alpha + \beta(b_1B_1 + b_2B_2 + b_3B_3) = \alpha + \beta_1B_1 + \beta_2B_2 + \beta_3B_3, \beta_i = \beta b_i, \alpha^2 + \beta^2 = 1,$$
$$b_1^2 + b_2^2 + b_3^2 = 1$$

I will use notation $so(\alpha, \beta, S)$ for such elements. They can be considered as points on unit radius sphere in the sense that any point $(\alpha, \beta_1, \beta_2, \beta_3) \in S^3$ parameterizes a $g$-qubit.

A pure qubit state in terms of conventional quantum mechanics is two dimensional unit value vector with complex components:

$$|\psi\rangle = \begin{pmatrix} z_1 \\ z_2 \end{pmatrix} = z_1\begin{pmatrix} 1 \\ 0 \end{pmatrix} + z_2\begin{pmatrix} 0 \\ 1 \end{pmatrix}, z_1^2 + z_2^2 = z_1\tilde{z}_1 + z_2\tilde{z}_2 = 1, z_k = z_k^1 + iz_k^2, k = 1,2$$

We can also think about such pure qubit states as points on $S^3$ because:

$$S^3 \ni \{z_1^1, z_1^2, z_2^1, z_2^2\} \leftarrow \{z = (z_1, z_2) \in C^2; |z|^2 = \tilde{z}_1z_1 + \tilde{z}_2z_2 = (z_1^1)^2 + (z_1^2)^2 + (z_2^1)^2 + (z_2^2)^2 = 1\} \quad (2.1)$$

Explicit relations, based on their parameterization by $S^3$ points, between elements $|\psi\rangle$ and elements $so(\alpha, \beta, S)$ can be established through the following construction[2].

Let basis bivector $B_1$ is chosen as defining the complex plane $S$, then we have (see multiplication rules (1.1)):

---

[2] Though both $so(\alpha, \beta, S)$ and $|\psi\rangle$ can be parametrized by points of $S^3$, it is not correct to say that $\{z = (z_1, z_2) \in C^2; |z|^2 = 1\}$ is sphere $S^3$ or $g$-qubit is sphere $S^3$.

$$so(\alpha,\beta,S) = (\alpha + \beta_1 B_1) + \beta_2 B_1 B_3 + \beta_3 B_3 = (\alpha + \beta_1 B_1) + (\beta_3 + \beta_2 B_1)B_3$$

Hence we get the map:

$$so(\alpha,\beta,S) \to |\psi\rangle = \begin{pmatrix} z_1^1 \\ z_2^1 \end{pmatrix} = \begin{pmatrix} \alpha + i\beta_1 \\ \beta_3 + i\beta_2 \end{pmatrix}, \text{ where } i \equiv B_1$$

In the similar way, for two other selections of complex plane we get:

$$so(\alpha,\beta,S) \to |\psi\rangle = \begin{pmatrix} z_1^2 \\ z_2^2 \end{pmatrix} = \begin{pmatrix} \alpha + i\beta_2 \\ \beta_1 + i\beta_3 \end{pmatrix}, i \equiv B_2,$$

and $$so(\alpha,\beta,S) \to |\psi\rangle = \begin{pmatrix} z_1^3 \\ z_2^3 \end{pmatrix} = \begin{pmatrix} \alpha + i\beta_3 \\ \beta_2 + i\beta_1 \end{pmatrix}, i \equiv B_3$$

So we have three different maps $so(\alpha,\beta,S) \xrightarrow{B_i} (z_1^i, z_2^i)$ defined by explicitly declared complex planes $B_i$ satisfying (1.1):

$$\left.\begin{array}{l} so(\alpha,\beta,S) = \alpha + \beta_1 B_1 + \beta_2 B_2 + \beta_3 B_3 = \\ \alpha + \beta_1 B_1 + (\beta_3 + \beta_2 B_1)B_3 = \\ \alpha + \beta_2 B_2 + (\beta_1 + \beta_3 B_2)B_1 = \\ \alpha + \beta_3 B_3 + (\beta_2 + \beta_1 B_3)B_2 \end{array}\right\} \Rightarrow \begin{cases} \begin{pmatrix} \alpha + i\beta_1 \\ \beta_3 + i\beta_2 \end{pmatrix}, i \leftarrow B_1 \\ \begin{pmatrix} \alpha + i\beta_2 \\ \beta_1 + i\beta_3 \end{pmatrix}, i \leftarrow B_2 \\ \begin{pmatrix} \alpha + i\beta_3 \\ \beta_2 + i\beta_1 \end{pmatrix}, i \leftarrow B_3 \end{cases} \quad (2.2)$$

There exists infinite number of options to select the triple $\{B_i\}$. It means that to recover a $g$-qubit $so(\alpha,\beta,S)$ in 3D associated with $|\psi\rangle = \begin{pmatrix} z_1 \\ z_2 \end{pmatrix}$ it is necessary, firstly, to define which bivector $B_{i_1}$ in 3D should be taken as defining "complex" plane and then to choose another bivector $B_{i_2}$, orthogonal to $B_{i_1}$. The third bivector $B_{i_3}$, orthogonal to both $B_{i_1}$ and $B_{i_2}$, is then defined by the first two by orientation (handedness, right screw in the used case ): $I_3 B_{i_1} I_3 B_{i_2} I_3 B_{i_3} = I_3$.

The conclusion is that to each single element (2.1) there corresponds infinite number of elements $so(\alpha,\beta,S)$ depending on choosing of a triple of orthonormal bivectors $\{B_1, B_2, B_3\}$ in 3D satisfying (1.1), and associating one of them with complex plane. This allows constructing map, fibration $G_3^+ \to C^2$ restricted to unit value elements.

## 3. Fiber bundle

Take general definition of fiber bundle as a set $(E, M, \pi, G, F)$ where $E$ is bundle (or total) space; $M$ - the base space; $F$ - standard fiber; $G$ - Lie group which acts effectively on $F$; $\pi$ - bundle projection: $\pi: E \to M$, such that each space $F_x \equiv \pi^{-1}(x)$, fiber at $x \in M$, is homeomorphic to standard fiber $F$.

We can think about the map $so(\alpha, \beta, S) \xrightarrow{B_i} (z_1^i, z_2^i)$ as a fiber bundle. In the current case, the fiber bundle will have $\{so(\alpha, \beta, S) = g \in G_3^+ : |g| = 1\}$ as total space and $\{z = (z_1, z_2) \in C^2; |z|^2 = 1\}$ as base space. I will denote them as $G_3^+|_{S^3}$ and $C^2|_{S^3}$ respectively. The projection $\pi: G_3^+|_{S^3} \to C^2|_{S^3}$ depends on which particular $B_i$ is taken from an arbitrary triple $\{B_1, B_2, B_3\}$ satisfying (1.1) as associated complex plane of complex vectors of $C^2$ and explicitly given by (2.2), so we should write $\pi: G_3^+|_{S^3} \xrightarrow{B_i} C^2|_{S^3}$.

By some reasons that will be explained a bit later I will use complex plane associated with $B_3$, so by (2.2) the projection is:

$$\pi: so(\alpha, \beta, S) = \alpha + \beta_1 B_1 + \beta_2 B_2 + \beta_3 B_3 \to \begin{pmatrix} \alpha + i\beta_3 \\ \beta_2 + i\beta_1 \end{pmatrix} \qquad (3.1)$$

Then for any $z = \begin{pmatrix} x_1 + iy_1 \\ x_2 + iy_2 \end{pmatrix} \in C^2|_{S^3}$ the fiber in $G_3^+|_{S^3}$ is comprised of all elements $F_z = x_1 + y_2 B_1 + x_2 B_2 + y_1 B_3$ with an arbitrary triple of orthonormal bivectors $\{B_1, B_2, B_3\}$ in 3D satisfying (1.1). That particularly means that standard fiber is equivalent to the group of rotations of the triple $\{y_2 B_1, x_2 B_2, y_1 B_3\}$ as a whole:

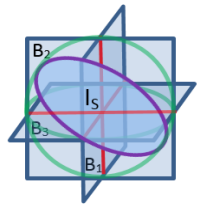 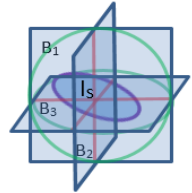

Fig 3a.                                Fig 3b.

Two example sections of a fiber. 3b received from 3a by 90 deg. counterclockwise rotation around vertical axis

All such rotations in $G_3^+$ are also identified by elements of $G_3^+\big|_{S^3}$ since for any bivector $B$ the result of its rotation is [3] (see, for example [6], [7]):

$$so(\gamma,\delta,S)\tilde{\ }Bso(\gamma,\delta,S), \quad \text{where } so(\gamma,\delta,S)\tilde{\ } = (\gamma+\delta_1 B_1+\delta_2 B_2+\delta_3 B_3)\tilde{\ } = \gamma-\delta_1 B_1-\delta_2 B_2-\delta_3 B_3$$

So, standard fiber is identified as $G_3^+\big|_{S^3}$ and composition of rotations is:

$$e^{-I s_2 \psi}\left(e^{-I s_1 \varphi} B e^{I s_1 \varphi}\right) e^{I s_2 \psi} = \left(e^{I s_1 \varphi} e^{I s_2 \psi}\right)\tilde{\ } B e^{I s_1 \varphi} e^{I s_2 \psi}$$

All that means that the fibration $G_3^+\big|_{S^3} \xrightarrow{B_3} C^2\big|_{S^3}$ is principal fiber bundle with standard fiber $G_3^+\big|_{S^3}$ and the group acting on it is the group of (right) multiplications by elements of $G_3^+\big|_{S^3}$.

Now the explanation why $B_3$ was taken as complex plane. As was shown in [1], [2] the variant of classical Hopf fibration

$$C^2\big|_{S^3} \to S^2 : \Pr(z_1, z_2) = \left(\tilde{z}_1 z_2 + z_1 \tilde{z}_2, i(\tilde{z}_1 z_2 - z_1 \tilde{z}_2), |z_1|^2 - |z_2|^2\right)$$

can be received as one of the cases of generalized Hopf fibrations:

$$G_3^+\big|_{S^3} \xrightarrow{B} S^2 : so(\alpha,\beta,S) \xrightarrow{B} so(\alpha,\beta,S)\tilde{\ } B so(\alpha,\beta,S) \to S^2.$$

The basis bivector $B_1$ gives:

$$so(\alpha,\beta,S) \xrightarrow{B_1} (\alpha - \beta I_S) B_1 (\alpha + \beta I_S) = (\alpha - \beta_1 B_1 - \beta_2 B_2 - \beta_3 B_3) B_1 (\alpha + \beta_1 B_1 + \beta_2 B_2 + \beta_3 B_3) =$$
$$\left[(\alpha^2 + \beta_1^2) - (\beta_2^2 + \beta_3^2)\right] B_1 + 2(\alpha\beta_3 + \beta_1\beta_2) B_2 + 2(\beta_1\beta_3 - \alpha\beta_2) B_3 \to$$
$$\left(\left[(\alpha^2 + \beta_1^2) - (\beta_2^2 + \beta_3^2)\right], 2(\alpha\beta_3 + \beta_1\beta_2), 2(\beta_1\beta_3 - \alpha\beta_2)\right) \quad (3.2)$$

and for two other basis bivectors:

$$so(\alpha,\beta,S) \xrightarrow{B_2} 2(\beta_1\beta_2 - \alpha\beta_3) B_1 + \left[(\alpha^2 + \beta_2^2) - (\beta_1^2 + \beta_3^2)\right] B_2 + 2(\alpha\beta_1 + \beta_2\beta_3) B_3 \to$$
$$\left(2(\beta_1\beta_2 - \alpha\beta_3), \left[(\alpha^2 + \beta_2^2) - (\beta_1^2 + \beta_3^2)\right], 2(\alpha\beta_1 + \beta_2\beta_3)\right) \quad (3.3)$$

$$so(\alpha,\beta,S) \xrightarrow{B_3} 2(\alpha\beta_2 + \beta_1\beta_3) B_1 + 2(\beta_2\beta_3 - \alpha\beta_1) B_2 + \left[(\alpha^2 + \beta_3^2) - (\beta_1^2 + \beta_2^2)\right] B_3 \to$$
$$\left(2(\alpha\beta_2 + \beta_1\beta_3), 2(\beta_2\beta_3 - \alpha\beta_1), \left[(\alpha^2 + \beta_3^2) - (\beta_1^2 + \beta_2^2)\right]\right) \quad (3.4)$$

---

[3] It is often convenient to write elements $so(\alpha,\beta,S)$ as exponents: $so(\alpha,\beta,S) = e^{I s \varphi}$.

In the literature mostly the third or the first variants are called Hopf fibrations. For my considerations it does not matter which variant to choose. I take the case (3.4) with $B_3$ as complex plane:

$$so(\alpha, \beta, S) = \alpha + \beta_1 B_1 + \beta_2 B_2 + \beta_3 B_3 = \alpha + \beta_3 B_3 + (\beta_2 + \beta_1 B_3) B_2 \to \begin{pmatrix} \alpha + i\beta_3 \\ \beta_2 + i\beta_1 \end{pmatrix}, i \leftarrow B_3$$

## 4. Tangent spaces.

We need to temporarily get back to the case of $G_2$ - geometric algebra on a plane [4].

Let an orthonormal basis $\{e_1, e_2\}$ is taken. It generates $G_2$ basis $\{1, e_1, e_2, e_1 e_2\}$ where $e_1 e_2 = e_1 \cdot e_2 + e_1 \wedge e_2$ is usual geometric product of two vectors: first member is scalar product of the vectors and second member is oriented area (bivector) swept by rotating $e_1$ to $e_2$ by the angle which is less than $\pi$.

The $G_2$ basis vectors satisfy particularly the properties:

$I_2^2 \equiv (e_1 e_2)^2 = -1$, $I_2 e_1 = -e_2$ (clockwise rotation), $e_1 I_2 = e_2$ (counterclockwise rotation), $I_2 e_2 = e_1$, $e_2 I_2 = -e_1$. I am using notation $I_2$ for unit bivector $e_1 e_2$.

For further convenience, let's construct a matrix basis isomorphic to $\{1, e_1, e_2, e_1 e_2\}$. Commonly used agreement will be that scalars are identified with scalar matrices, for example: $\alpha \equiv \begin{pmatrix} \alpha & 0 \\ 0 & \alpha \end{pmatrix}$.

For the $G_2$ case the second order matrices will suffice to get necessary isomorphism. Let's take

$\varepsilon_1 = \begin{pmatrix} 1 & 0 \\ 0 & -1 \end{pmatrix}$ and $\varepsilon_2 = \begin{pmatrix} 0 & 1 \\ 1 & 0 \end{pmatrix}$ [4] as corresponding to geometric basis vectors $e_1$ and $e_2$. They satisfy the properties mentioned above, for example: $\varepsilon_i^2 = \begin{pmatrix} 1 & 0 \\ 0 & 1 \end{pmatrix}$, $I_2^2 = (\varepsilon_1 \varepsilon_2)^2 = \begin{pmatrix} 0 & 1 \\ -1 & 0 \end{pmatrix}^2 = \begin{pmatrix} -1 & 0 \\ 0 & -1 \end{pmatrix}$,

$I_2 \varepsilon_1 = \begin{pmatrix} 0 & 1 \\ -1 & 0 \end{pmatrix}\begin{pmatrix} 1 & 0 \\ 0 & -1 \end{pmatrix} = \begin{pmatrix} 0 & -1 \\ -1 & 0 \end{pmatrix} = -\varepsilon_2$, etc.

The operation representing scalar product gives:

---

[4] I begin with ordinary real component matrices. That is why the order of Pauli matrices in my following definition is different from usual one. An analogue of "imaginary" unit should logically be the last one, after pure real items.

$$\varepsilon_1 \cdot \varepsilon_2 = \frac{1}{2}(\varepsilon_1 \varepsilon_2 + \varepsilon_2 \varepsilon_1) = \frac{1}{2}\left(\begin{pmatrix} 0 & 1 \\ -1 & 0 \end{pmatrix} + \begin{pmatrix} 0 & -1 \\ 1 & 0 \end{pmatrix}\right) = \begin{pmatrix} 0 & 0 \\ 0 & 0 \end{pmatrix} \quad (4.1)$$

as it should be. If we take two arbitrary vectors expanded in basis $\{\varepsilon_1, \varepsilon_2\}$:

$$\vec{a} = \alpha_1 \varepsilon_1 + \alpha_2 \varepsilon_2 = \begin{pmatrix} \alpha_1 & 0 \\ 0 & -\alpha_1 \end{pmatrix} + \begin{pmatrix} 0 & \alpha_2 \\ \alpha_2 & 0 \end{pmatrix} = \begin{pmatrix} \alpha_1 & \alpha_2 \\ \alpha_2 & -\alpha_1 \end{pmatrix}$$

$$\vec{b} = \beta_1 \varepsilon_1 + \beta_2 \varepsilon_2 = \begin{pmatrix} \beta_1 & 0 \\ 0 & -\beta_1 \end{pmatrix} + \begin{pmatrix} 0 & \beta_2 \\ \beta_2 & 0 \end{pmatrix} = \begin{pmatrix} \beta_1 & \beta_2 \\ \beta_2 & -\beta_1 \end{pmatrix}$$

then their product is:

$$\vec{a}\vec{b} = \begin{pmatrix} \alpha_1 \beta_1 + \alpha_2 \beta_2 & 0 \\ 0 & \alpha_1 \beta_1 + \alpha_2 \beta_2 \end{pmatrix} + \begin{pmatrix} 0 & -\alpha_2 \beta_1 + \alpha_1 \beta_2 \\ -\alpha_1 \beta_2 + \alpha_2 \beta_1 & 0 \end{pmatrix} = (\vec{a} \cdot \vec{b})\begin{pmatrix} 1 & 0 \\ 0 & 1 \end{pmatrix} + (\alpha_1 \beta_2 - \alpha_2 \beta_1)I_2$$

The first member is usual scalar product and the second one is bivector of the value equal to the area of parallelogram with the sides $\vec{a}$ and $\vec{b}$.

Important thing to keep in mind is that multiplication of any vector by unit bivector $I_2$ rotates the vector by $\pm\frac{\pi}{2}$:

$$\vec{a}I_2 = \begin{pmatrix} \alpha_1 & \alpha_2 \\ \alpha_2 & -\alpha_1 \end{pmatrix}\begin{pmatrix} 0 & 1 \\ -1 & 0 \end{pmatrix} = \begin{pmatrix} -\alpha_2 & \alpha_1 \\ \alpha_1 & \alpha_2 \end{pmatrix} = -\alpha_2 \varepsilon_1 + \alpha_1 \varepsilon_2 \quad \text{(counterclockwise rotation)}$$

$$I_2\vec{a} = \begin{pmatrix} 0 & 1 \\ -1 & 0 \end{pmatrix}\begin{pmatrix} \alpha_1 & \alpha_2 \\ \alpha_2 & -\alpha_1 \end{pmatrix} = \begin{pmatrix} \alpha_2 & -\alpha_1 \\ -\alpha_1 & -\alpha_2 \end{pmatrix} = \alpha_2 \varepsilon_1 - \alpha_1 \varepsilon_2 \quad \text{(clockwise rotation)}$$

This remains valid for any unit bivector of the same orientation as $I_2$. We can conclude that such multiplications give basis vectors of the tangent spaces to the original vectors.

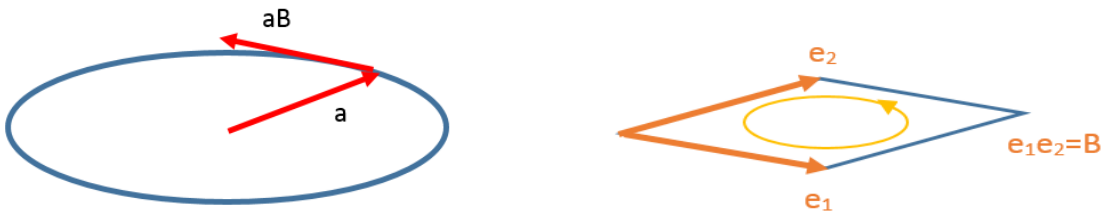

This is identical to considering even elements $\alpha_1 + B\alpha_2$ corresponding to vectors and their multiplication by $B$: $(\alpha_1 + B\alpha_2)B = -\alpha_2 + B\alpha_1$. Elements $\alpha_1 + B\alpha_2$ and $-\alpha_2 + B\alpha_1$ are orthogonal: $\langle \alpha_1 + B\alpha_2, -\alpha_2 + B\alpha_1 \rangle_0 = (\alpha_1 + B\alpha_2)\cdot(-\alpha_2 - B\alpha_1)_0 = 0$ (index 0 means scalar part).

General element of $G_2$ in matrix basis $\{1, \varepsilon_1, \varepsilon_2, \varepsilon_1\varepsilon_2 \equiv I_2\}$, isomorphic to geometrical basis $\{1, e_1, e_2, e_1 e_2\}$, is:

$$g_2 = \alpha_0 \begin{pmatrix} 1 & 0 \\ 0 & 1 \end{pmatrix} + \alpha_1 \begin{pmatrix} 1 & 0 \\ 0 & -1 \end{pmatrix} + \alpha_2 \begin{pmatrix} 0 & 1 \\ 1 & 0 \end{pmatrix} + \alpha_3 \begin{pmatrix} 0 & 1 \\ -1 & 0 \end{pmatrix} = \begin{pmatrix} \alpha_0 + \alpha_1 & \alpha_2 + \alpha_3 \\ \alpha_2 - \alpha_3 & \alpha_0 - \alpha_1 \end{pmatrix}$$

This is arbitrary real valued matrix of the second order. Inversely, any matrix of second order can be uniquely mapped to the element of $G_2$ in the basis $\{1, \varepsilon_1, \varepsilon_2, \varepsilon_1\varepsilon_2 \equiv I_2\}$:

$$\begin{pmatrix} x & y \\ z & t \end{pmatrix} \Rightarrow \frac{x+t}{2} + \frac{x-t}{2}\varepsilon_1 + \frac{y+z}{2}\varepsilon_2 + \frac{y-z}{2}I_2$$

To upgrade to $G_3$, we need basis matrices of a higher order because a second order non-zero matrix orthogonal in the sense of scalar product (4.1) to both $\varepsilon_1$ and $\varepsilon_2$ does not exist.

It is easy to verify that the three, playing the role of three-dimensional orthonormal basis, necessary matrices can be taken as 4th-order block matrices:

$$\sigma_1 = \begin{pmatrix} I & 0 \\ 0 & -I \end{pmatrix}, \sigma_2 = \begin{pmatrix} 0 & I \\ I & 0 \end{pmatrix}, \sigma_3 = \begin{pmatrix} 0 & -I_2 \\ I_2 & 0 \end{pmatrix},$$

where $I = \begin{pmatrix} 1 & 0 \\ 0 & 1 \end{pmatrix}$ - scalar matrix corresponding to scalar 1, $I_2 = \begin{pmatrix} 0 & 1 \\ -1 & 0 \end{pmatrix} = \varepsilon_1\varepsilon_2$. Easy to see that $\sigma_1$ and $\sigma_2$ are received from $\varepsilon_1$ and $\varepsilon_2$ by replacing scalars by corresponding scalar matrices.

By multiplications in full $4\times 4$ form one can easily proof that block-wise multiplications are correct, the basis $\{\sigma_1, \sigma_2, \sigma_3\}$ is orthonormal, anticommutative, and $\sigma_2\sigma_3$, $\sigma_3\sigma_1$ and $\sigma_1\sigma_2$ satisfy requirements (1.1).

Taking $I_3 = \sigma_1\sigma_2\sigma_3 = \begin{pmatrix} I_2 & 0 \\ 0 & I_2 \end{pmatrix}$, multiplications give: $\sigma_2\sigma_3 = \begin{pmatrix} I_2 & 0 \\ 0 & -I_2 \end{pmatrix} = I_3 \begin{pmatrix} I & 0 \\ 0 & -I \end{pmatrix} = I_3\sigma_1$,

$\sigma_3\sigma_1 = \begin{pmatrix} 0 & I_2 \\ I_2 & 0 \end{pmatrix} = I_3 \begin{pmatrix} 0 & I \\ I & 0 \end{pmatrix} = I_3\sigma_2$, $\sigma_1\sigma_2 = \begin{pmatrix} 0 & I \\ -I & 0 \end{pmatrix} = I_3 \begin{pmatrix} 0 & -I_2 \\ I_2 & 0 \end{pmatrix} = I_3\sigma_3$. Then, since $I_3^2 = -1$,

we easily get:

$$(\sigma_2\sigma_3)(\sigma_3\sigma_1) = -\sigma_1\sigma_2, \quad (\sigma_2\sigma_3)(\sigma_1\sigma_2) = \sigma_3\sigma_1, \quad (\sigma_3\sigma_1)(\sigma_1\sigma_2) = -\sigma_2\sigma_3$$

that is exactly (1.1) with basis bivectors $B_1 = \sigma_2\sigma_3$, $B_2 = \sigma_3\sigma_1$ and $B_3 = \sigma_1\sigma_2$.

Following all that we can write general element $g_3$ of algebra $G_3$ expanded in formal matrix basis $\{1, \sigma_1, \sigma_2, \sigma_3, \sigma_2\sigma_3, \sigma_3\sigma_1, \sigma_1\sigma_2, \sigma_1\sigma_2\sigma_3\} \equiv \{1, \sigma_1, \sigma_2, \sigma_3, I_3\sigma_1, I_3\sigma_2, I_3\sigma_3, I_3\}$ as:

$$g_3 = \alpha I + \alpha_1\sigma_1 + \alpha_2\sigma_2 + \alpha_3\sigma_3 + \beta_1\sigma_2\sigma_3 + \beta_2\sigma_3\sigma_1 + \beta_3\sigma_1\sigma_2 + \beta_0\sigma_1\sigma_2\sigma_3 =$$

$$\begin{pmatrix} (\alpha+\alpha_1)+I_2(\beta_0+\beta_1) & (\alpha_2+\beta_3)+I_2(\beta_2-\alpha_3) \\ (\alpha_2-\beta_3)+I_2(\beta_2+\alpha_3) & (\alpha-\alpha_1)+I_2(\beta_0-\beta_1) \end{pmatrix} \quad (4.2)$$

This is arbitrary "complex" valued matrix where "imaginary unit" is matrix $I_2 = \begin{pmatrix} 0 & I \\ -I & 0 \end{pmatrix}$, corresponding to bivector expanded in the first two basis vectors $\sigma_1$, $\sigma_2$, that geometrically is bivector $B_3$. Inversely, any "complex" valued matrix can be written as $g_3$ element in matrix basis $\{1, \sigma_1, \sigma_2, \sigma_3, I_3\sigma_1, I_3\sigma_2, I_3\sigma_3, I_3\}$ isomorphic to geometrical basis $\{1, e_1, e_2, e_3, e_2e_3, e_3e_1, e_1e_2, e_1e_2e_3\}$:

$$\begin{pmatrix} s+I_2t & u+I_2v \\ x+I_2y & z+I_2w \end{pmatrix} = \frac{s+z}{2} + \frac{s-z}{2}\sigma_1 + \frac{u+x}{2}\sigma_2 + \frac{y-v}{2}\sigma_3 + \frac{t-w}{2}\sigma_2\sigma_3 + \frac{y+v}{2}\sigma_3\sigma_1 + \frac{u-x}{2}\sigma_1\sigma_2 + \frac{t+w}{2}I_3$$

If $g_3$ belongs to even subalgebra $G_3^+$ then, by identifying $\sigma_2\sigma_3 \Leftrightarrow B_1$, $\sigma_3\sigma_1 \Leftrightarrow B_2$, $\sigma_1\sigma_2 \Leftrightarrow B_3$, we have from (4.2) the correspondence:

$$g_3^+ = \alpha + \beta_1 B_1 + \beta_2 B_2 + \beta_3 B_3 \Leftrightarrow \begin{pmatrix} \alpha + I_2\beta_1 & \beta_3 + I_2\beta_2 \\ -\beta_3 + I_2\beta_2 & \alpha - I_2\beta_1 \end{pmatrix} \quad (4.3)$$

In the same way as in the $G_2$ case, multiplications of $so(\alpha, \beta_1, \beta_2, \beta_3) = \alpha + \beta_1 B_1 + \beta_2 B_2 + \beta_3 B_3$ by basis bivectors $B_i$ give basis bivectors of tangent spaces to original bivectors:

$$T_1 = so(\alpha, \beta_1, \beta_2, \beta_3)B_1 = (\alpha + \beta_1 B_1 + \beta_2 B_2 + \beta_3 B_3)B_1 = -\beta_1 + \alpha B_1 - \beta_3 B_2 + \beta_2 B_3$$

$$T_2 = so(\alpha, \beta_1, \beta_2, \beta_3)B_2 = (\alpha + \beta_1 B_1 + \beta_2 B_2 + \beta_3 B_3)B_2 = -\beta_2 + \beta_3 B_1 + \alpha B_2 - \beta_1 B_3$$

$$T_3 = so(\alpha, \beta_1, \beta_2, \beta_3)B_3 = (\alpha + \beta_1 B_1 + \beta_2 B_2 + \beta_3 B_3)B_3 = -\beta_3 - \beta_2 B_1 + \beta_1 B_2 + \alpha B_3$$

These $G_3^+|_{S^3}$ elements are orthogonal to $\alpha + \beta_1 B_1 + \beta_2 B_2 + \beta_3 B_3$ and to each other, and are the tangent space basis elements at points $\alpha + \beta_1 B_1 + \beta_2 B_2 + \beta_3 B_3$.

Projections of $T_i$ onto $C^2|_{S^3}$ are:

$$\pi(T_1) = \begin{pmatrix} -\beta_1 + i\beta_2 \\ -\beta_3 + i\alpha \end{pmatrix}, \quad \pi(T_2) = \begin{pmatrix} -\beta_2 - i\beta_1 \\ \alpha + i\beta_3 \end{pmatrix}, \quad \pi(T_3) = \begin{pmatrix} -\beta_3 + i\alpha \\ \beta_1 - i\beta_2 \end{pmatrix}$$

These elements of $C^2|_{S^3}$ are orthogonal, in the sense of Euclidean scalar product in $C^2$:

$\langle z, w \rangle = \mathrm{Re}(\tilde{z}_1 w_1 + \tilde{z}_2 w_2)$, to each other and to the projection $\pi(so(\alpha, \beta_1, \beta_2, \beta_3)) = \begin{pmatrix} \alpha + i\beta_3 \\ \beta_2 + i\beta_1 \end{pmatrix}$ of

the original state in $G_3^+|_{S^3}$. They are the tangent space basis elements in $C^2|_{S^3}$ at points

$$\begin{pmatrix} z_1 \\ z_2 \end{pmatrix} = \begin{pmatrix} \alpha + i\beta_3 \\ \beta_2 + i\beta_1 \end{pmatrix}.$$

## 5. Hamiltonians in $C^2|_{S^3}$ and their lifts in $G_3$

Take self-adjoint (Hamiltonian) linear operator in $C^2$ in its most general form:

$H = aI + b\sigma_1 + c\sigma_2 + d\sigma_3 = \begin{pmatrix} a+b & c - I_2 d \\ c + I_2 d & a - b \end{pmatrix}$. Its corresponding $G_3$ element, see (4.3), is

$a + I_3(bB_1 + cB_2 + dB_3)$ and does not belong to $G_3^+|_{S^3}$, though the result of its action on an element

from $C^2|_{S^3}$ (qubit), has $\pi^{-1}$ preimage in $G_3^+|_{S^3}$. Let's prove that. Take a qubit and a Hamiltonian in $C^2|_{S^3}$:

$$z = \begin{pmatrix} x_1 + iy_1 \\ x_2 + iy_2 \end{pmatrix} = \begin{pmatrix} r_1 e^{i\psi_1} \\ r_2 e^{i\psi_2} \end{pmatrix}, \quad r_k = \sqrt{x_k^2 + y_k^2}, \quad \cos\psi_k = \frac{x_k}{\sqrt{x_k^2 + y_k^2}}, \quad \sin\psi_k = \frac{y_k}{\sqrt{x_k^2 + y_k^2}}, \quad k = 1,2$$

$$H = \begin{pmatrix} a+b & Re^{i\psi} \\ Re^{-i\psi} & a-b \end{pmatrix}, \quad R = \sqrt{c^2 + d^2}, \quad \cos\psi = \frac{c}{\sqrt{c^2 + d^2}}, \quad \sin\psi = \frac{d}{\sqrt{c^2 + d^2}}$$

Then:

$$Hz = \begin{pmatrix} a+b & Re^{i\psi} \\ Re^{-i\psi} & a-b \end{pmatrix} \begin{pmatrix} r_1 e^{i\psi_1} \\ r_2 e^{i\psi_2} \end{pmatrix} =$$
$$\begin{pmatrix} (a+b)r_1 \cos\psi_1 + Rr_2 \cos(\psi + \psi_2) + i((a+b)r_1 \sin\psi_1 + Rr_2 \sin(\psi + \psi_2)) \\ (a-b)r_2 \cos\psi_2 + Rr_1 \cos(\psi - \psi_1) + i((a-b)r_2 \sin\psi_2 - Rr_1 \sin(\psi - \psi_1)) \end{pmatrix}$$

(5.1)

For each single qubit $z = \begin{pmatrix} x_1 + iy_1 \\ x_2 + iy_2 \end{pmatrix} \in C^2 |_{S^3}$ its fiber (full preimage) in $G_3^+ |_{S^3}$ is

$F_z = x_1 + y_2 B_1 + x_2 B_2 + y_1 B_3$ with an arbitrary triple of orthonormal bivectors $\{B_1, B_2, B_3\}$ in 3D satisfying (1.1). In exponential form:

$$F_z = x_1 + y_2 B_1 + x_2 B_2 + y_1 B_3 = x_1 + \sqrt{1-x_1^2}\left(\frac{y_2}{\sqrt{1-x_1^2}} B_1 + \frac{x_2}{\sqrt{1-x_1^2}} B_2 + \frac{y_1}{\sqrt{1-x_1^2}} B_3\right)$$

$$= \cos\varphi + \sin\varphi(b_1 B_1 + b_2 B_2 + b_3 B_3) = e^{I_S \varphi}, \qquad (5.2)$$

where

$$\left.\begin{array}{l} \cos\varphi = x_1, \sin\varphi = \sqrt{1-x_1^2} \\ b_1 = \dfrac{y_2}{\sqrt{1-x_1^2}}, b_2 = \dfrac{x_2}{\sqrt{1-x_1^2}}, b_3 = \dfrac{y_1}{\sqrt{1-x_1^2}} \\ I_S = b_1 B_1 + b_2 B_2 + b_3 B_3 \end{array}\right\} \qquad (5.3)$$

Using (5.2) we get from (5.3) the fiber at $Hz$:

$$F_{Hz} = e^{I_{S(H)} \varphi(H)}, \qquad (5.4)$$

where

$$\left.\begin{array}{l} \cos\varphi(H) = (a+b)r_1 \cos\psi_1 + Rr_2 \cos(\psi+\psi_2); \\ I_{S(H)} = b_1(H) B_1 + b_2(H) B_2 + b_3(H) B_3; \\ b_1(H) = \dfrac{(a-b)r_2 \sin\psi_2 - Rr_1 \sin(\psi-\psi_1)}{\sqrt{1-\cos^2\varphi(H)}}, b_2(H) = \dfrac{(a-b)r_2 \cos\psi_2 + Rr_1 \cos(\psi-\psi_1)}{\sqrt{1-\cos^2\varphi(H)}}, \\ b_3(H) = \dfrac{(a+b)r_1 \sin\psi_1 + Rr_2 \sin(\psi+\psi_2)}{\sqrt{1-\cos^2\varphi(H)}} \end{array}\right\} \qquad (5.5)$$

We can also explicitly write the element $g_3(H) \in G_3^+ |_{S^3}$ which, acting (from the right) on $\pi^{-1}(z)$ gives $\pi^{-1}(Hz)$:

$$\pi^{-1}(z) g_3(H) = \pi^{-1}(Hz)$$

From (5.1) and (5.4):

$$e^{I_S \varphi} g_3(H) = e^{I_{S(H)} \varphi(H)} \Rightarrow g_3(H) = e^{-I_S \varphi} e^{I_{S(H)} \varphi(H)}$$

where $I_S$, $\varphi$, $I_{S(H)}$ and $\varphi(H)$ are defined by (5.3) and (5.5).

## 6. Clifford translations

Suppose, action on $z$ in $C^2|_{S^3}$ is multiplication by an exponent: $Cl_\psi(z) = e^{i\psi} z$, often called Clifford translation[5], that's:

$$Cl_\psi(z) = e^{i\psi} \begin{pmatrix} r_1 e^{i\psi_1} \\ r_2 e^{i\psi_2} \end{pmatrix} = \begin{pmatrix} r_1 \cos(\psi + \psi_1) + i r_1 \sin(\psi + \psi_1) \\ r_2 \cos(\psi + \psi_2) + i r_2 \sin(\psi + \psi_2) \end{pmatrix}$$

Then the fiber of $Cl_\psi(z)$ in $G_3^+|_{S^3}$ is:

$$\begin{aligned} F_{Cl_\psi(z)} &= r_1 \cos(\psi + \psi_1) + r_2 \sin(\psi + \psi_2) B_1 + r_2 \cos(\psi + \psi_2) B_2 + r_1 \sin(\psi + \psi_1) B_3 = \\ &r_1 \cos(\psi + \psi_1) + r_1 \sin(\psi + \psi_1) B_3 + [r_2 \cos(\psi + \psi_2) + r_2 \sin(\psi + \psi_2) B_3] B_2 = \\ &r_1 e^{B_3 \psi} e^{B_3 \psi_1} + r_2 e^{B_3 \psi} e^{B_3 \psi_2} B_2 = e^{B_3 \psi} F_z \end{aligned} \quad (6.1)$$

In other words, Clifford translations in $C^2|_{S^3}$ are equivalent to multiplications of fibers in $G_3^+|_{S^3}$ by standard fiber elements with bivector part $B_3$ (which is associated with formal imaginary unit $i$) and the same phase $\psi$.

In commonly used quantum mechanics all states from the set $Cl_\psi(z)$ are considered as identical to the state $z$, they have the same values of observables. That is not commonly true for $g$-qubits. Only in the case when all the observables common plane, say $I_S$, (unspecified in the $C^2$ model[6]) is the same as the plane associated with complex plane in $Cl_\psi(z)$, the $g$-qubit $e^{I_S \psi} F_z$ leaves any such observable $e^{I_S \varphi}$ unchanged:

$$\tilde{F}_z e^{-I_S \psi} e^{I_S \varphi} e^{I_S \psi} F_z = \tilde{F}_z e^{I_S \varphi} F_z$$

Suppose angle $\psi$ in $Cl_\psi(z)$ is varying. Then the tangent to Clifford orbit is

$$\frac{\partial}{\partial \psi} Cl_\psi(z) = \frac{\partial}{\partial \psi} e^{i\psi} z = i e^{i\psi} z = i Cl_\psi(z).$$

---

[5] It is called "translation" because does not change distance between elements.
[6] In the $G_3^+$ model observables are also elements of $G_3^+$, see [1], [2]

It is orthogonal to $Cl_\psi(z)$ in the sense of usual scalar product in $C^2$: $\langle u, v \rangle = \text{Re}(\tilde{u}_1 v_1 + \tilde{u}_2 v_2)$, and remains on $C^2|_{S^3}$: $|iCl_\psi(z)| = 1$ (translational velocity of Clifford translation is one).

For any $z = \begin{pmatrix} z_1 \\ z_2 \end{pmatrix} \in C^2|_{S^3}$ define $w = \begin{pmatrix} \tilde{z}_2 \\ -\tilde{z}_1 \end{pmatrix} \in C^2|_{S^3}$. The elements $iz$, $w$ and $iw$ are orthogonal to $z$ and pairwise orthogonal. That means they span tangent space and the plane spanned by $\{w, iw\}$ is orthogonal to the Clifford orbit plane spanned by $\{z, iz\}$.

The vector of translational velocity rotates while moving in the orbit plane. Both $w$ and $iw$ also rotates with the same unit value angular velocity since $\frac{\partial}{\partial \psi} Cl_\psi(w) = \frac{\partial}{\partial \psi} e^{i\psi} w = e^{i\psi} iw = Cl_\psi(iw)$ and

$$\frac{\partial}{\partial \psi} Cl_\psi(iw) = \frac{\partial}{\partial \psi} e^{i\psi} iw = ie^{i\psi} iw = -Cl_\psi(w).$$

To make the planes of rotations clearly identified, that is difficult to do with formal imaginary unit, let's lift Clifford translation to $G_3^+|_{S^3}$ using (6.1). Translational velocity, similar to the $C^2|_{S^3}$ case, is

$$\frac{\partial}{\partial \psi} F_{Cl_\psi(z)} = \frac{\partial}{\partial \psi} \left( e^{B_3 \psi} F_z \right) = B_3 F_{Cl_\psi(z)} \text{ and is orthogonal to } F_{Cl_\psi(z)}:$$

$$\left\langle F_{Cl_\psi(z)}, B_3 F_{Cl_\psi(z)} \right\rangle_0 = \left( F_{Cl_\psi(z)} \tilde{F}_{Cl_\psi(z)} \tilde{B}_3 \right)_0 = (B_3)_0 = 0$$

Two other components of the tangent space, orthogonal to $F_{Cl_\psi(z)}$ and $B_3 F_{Cl_\psi(z)}$ at any point of the orbit, are $B_1 F_{Cl_\psi(z)}$ and $B_2 F_{Cl_\psi(z)}$[7]. Their velocities while moving along Clifford orbit are:

$$\frac{\partial}{\partial \psi} \left( B_1 F_{Cl_\psi(z)} \right) = \frac{\partial}{\partial \psi} \left( B_1 e^{B_3 \psi} F_z \right) = B_1 B_3 F_{Cl_\psi(z)} = B_2 F_{Cl_\psi(z)} \text{ (derivative of } B_1 F_{Cl_\psi(z)} \text{ is orthogonal to}$$

$$B_1 F_{Cl_\psi(z)} \text{ and looking in the direction } B_2 F_{Cl_\psi(z)})$$

$$\frac{\partial}{\partial \psi} \left( B_2 F_{Cl_\psi(z)} \right) = \frac{\partial}{\partial \psi} \left( B_2 e^{B_3 \psi} F_z \right) = B_2 B_3 F_{Cl_\psi(z)} = -B_1 F_{Cl_\psi(z)} \text{ (derivative of } B_2 F_{Cl_\psi(z)} \text{ is orthogonal to}$$

$$B_2 F_{Cl_\psi(z)} \text{ and looking in the direction } -B_1 F_{Cl_\psi(z)})$$

These two equations explicitly show that the two tangents, orthogonal to Clifford translation velocity, rotate in moving plane $\{B_1 F_{Cl_\psi(z)}, B_2 F_{Cl_\psi(z)}\}$ with the same unit value rotational velocity. Interesting to notice that the triple of the translational velocity and two rotational velocities has orientation opposite

---

[7] Clearly, the three $B_i F_{Cl_\psi(z)}$ are identical to earlier considered tangents $T_i$

to the triple of tangents: if the tangents $\{B_1 F_{Cl_\psi(z)}, B_2 F_{Cl_\psi(z)}, B_3 F_{Cl_\psi(z)}\}$ have right screw orientation, the speed triple $\{-B_1 F_{Cl_\psi(z)}, B_2 F_{Cl_\psi(z)}, B_3 F_{Cl_\psi(z)}\}$ is left screw.

If a fiber, $g$-qubit, makes full circle in Clifford translation: $F_z \to F_{Cl_\psi(z)} = e^{B_3 \psi} F_z$, $0 \leq \psi \leq 2\pi$, both $B_1 F_{Cl_\psi(z)}$ and $B_2 F_{Cl_\psi(z)}$ also make full rotation in their own plane by $2\pi$. This is special case of the $g$-qubit Berry phase incrementing in a closed curve quantum state path.

## 7. Conclusions

Evolution of a quantum state described in terms of $G_3^+$ gives more detailed information about two state system compared to the $C^2$ Hilbert space model. It confirms the idea that distinctions between "quantum" and "classical" states become less deep if a more appropriate mathematical formalism is used. The paradigm spreads from trivial phenomena like tossed coin experiment [3] to recent results on entanglement and Bell theorem [8] where the former was demonstrated as not exclusively quantum property.

## Works Cited

[1]  A. Soiguine, "Geometric Algebra, Qubits, Geometric Evolution, and All That," January 2015. [Online]. Available: http://arxiv.org/abs/1502.02169.

[2]  A. Soiguine, "What quantum "state" really is?," June 2014. [Online]. Available: http://arxiv.org/abs/1406.3751.

[3]  A. Soiguine, "A tossed coin as quantum mechanical object," September 2013. [Online]. Available: http://arxiv.org/abs/1309.5002.

[4]  A. Soiguine, Vector Algebra in Applied Problems, Leningrad: Naval Academy, 1990 (in Russian).

[5]  A. Soiguine, "Complex Conjugation - Realtive to What?," in *Clifford Algebras with Numeric and Symbolic Computations*, Boston, Birkhauser, 1996, pp. 285-294.

[6]  D. Hestenes, New Foundations of Classical Mechanics, Dordrecht/Boston/London: Kluwer Academic Publishers, 1999.

[7] C. Doran and A. Lasenby, Geometric Algebra for Physicists, Cambridge: Cambridge University Press, 2010.

[8] X.-F. Qian, B. Little, J. C. Howell and J. H. Eberly, "Shifting the quantum-classical boundary: theory and experiment for statistically classical optical fields," *Optica,* pp. 611-615, 20 July 2015.

[9] J. W. Arthur, Understanding Geometric Algebra for Electromagnetic Theory, John Wiley & Sons, 2011.

[10] D. Chruscinski and A. Jamiolkowski, Geometric Phases in Classical and Quantum Mechanics, Boston: Birkhauser, 2004.